\documentclass{acm_sen_article}

\usepackage{cite}
\usepackage{amsmath,amssymb,amsfonts}
\usepackage{algorithmic}
\usepackage{graphicx}
\usepackage{textcomp}
\usepackage{graphicx}
\usepackage[table,x11names]{xcolor}
\usepackage{framed}
\usepackage{csquotes}
\usepackage{dirtytalk}
\usepackage{lscape}
\usepackage{comment}
\PassOptionsToPackage{hyphens}{url}
\usepackage{hyperref}
\usepackage{flushend}

\begin{document}

\title{Guidelines for Artifacts to Support Industry-Relevant Research on Self-Adaptation}


\author{Danny Weyns$^{1,2}$, 
Ilias Gerostathopoulos$^{3}$, 
Barbora Buhnova$^{4}$, 
Nicol\'as Cardozo$^{5}$, 
Emilia Cioroaica$^{6}$, \\
Ivana Dusparic$^{7}$,
Lars Grunske$^{8}$, 
Pooyan Jamshidi$^{9}$, 
Christine Julien$^{10}$,
Judith Michael$^{11}$, \\
Gabriel Moreno$^{12}$, 
Shiva Nejati$^{13}$, 
Patrizio Pelliccione$^{14}$, 
Federico Quin$^{1}$, 
Genaina Rodrigues$^{15}$, \\
Bradley Schmerl$^{12}$,
Marco Vieira$^{16}$, 
Thomas Vogel$^{8,17}$, 
Rebekka Wohlrab$^{12}$
 \\
\\
$^{1}$Katholieke Universiteit Leuven; 
$^{2}$Linnaeus University;  
$^{3}$Vrije Universiteit Amsterdam; \\
$^{4}$Masaryk University; 
$^{5}$Universidad de los Andes; 
$^{6}$Fraunhofer IESE; 
$^{7}$Trinity College Dublin; \\
$^{8}$Humboldt-Universit\"at zu Berlin; 
$^{9}$University of South Carolina; 
$^{10}$University of Texas Austin; \\
$^{11}$RWTH Aachen University; 
$^{12}$Carnegie Mellon University; 
$^{13}$University of Ottawa; \\
$^{14}$Gran Sasso Science Institute; 
$^{15}$Universidade de Bras\'ilia; \\
$^{16}$University of Coimbra; 
$^{17}$Paderborn University \\
\email{danny.weyns@kuleuven.be, ilias.gerostathopoulos@vu.nl}
}

\maketitle

\begin{abstract}
Artifacts support evaluating new research results and help comparing them with the state of the art in a field of interest. Over the past years, several artifacts have been introduced to support research in the field of self-adaptive systems. While these artifacts have shown their value, it is not clear to what extent these artifacts support research on problems in self-adaptation that are relevant to industry. This paper provides a set of guidelines for artifacts that aim at supporting industry-relevant research on self-adaptation. The guidelines that are grounded on data obtained from a survey with practitioners were derived during working sessions at the 17th International Symposium on Software Engineering for Adaptive and Self-Managing Systems. Artifact providers can use the guidelines for aligning future artifacts with industry needs; they can also be used to evaluate the industrial relevance of existing artifacts. We also propose an artifact template. 
\end{abstract}

\section{Introduction}
The term \textit{artifact} has different meanings in the field of software engineering. In software development communities, artifacts refer to documents, deliverables, and work products~\cite{FernandezBVMBKW19}. Software engineering conferences increasingly use the term artifact in association with publications. Such artifacts can range from ``functional'' i.e., artifacts associated with the research that are documented and exercisable, and include appropriate evidence of verification and validation; to ``replicated'' i.e.,  the main results of the paper have been independently obtained in a subsequent study by a person or team other than the authors, without the use of author-supplied artifacts.\footnote{\url{https://conf.researchr.org/track/icse-2022/icse-2022-artifact-evaluation}} 
In this paper, we refer to an artifact as a tangible object that is created with the intention to drive research advances and compare and contrast alternative approaches.\footnote{\url{https://conf.researchr.org/track/seams-2022/seams-2022-papers\#Artifacts}} Our particular focus is on artifacts that support research in the field of self-adaptation and its applicability in practice. 

Generally, different types of artifacts can be provided. \textit{Testbeds} refer to implementations of systems (or parts of systems) that support the study and evaluation of open challenges in self-adaptive systems. \textit{Model problems} provide detailed specifications of systems that pose fundamental or characteristic challenges that self-adaptive systems should address, either in general or in specific domains. \textit{Frameworks} refer to concrete or conceptual platforms with generic functionality that can be selectively specialized by implementing self-adaptation techniques or algorithms that are potentially useful in different contexts and use. \textit{Datasets}  (e.g., logging data, sensor data, system traces, survey raw data)  can be used to develop, evaluate, and compare self-adaptation approaches. These common types of artifacts are representative classes of the existing artifact landscape. 

Over the past decade, researchers in the field of self-adaptation developed about 30 artifacts~\cite{SEAMSartifacts}. A pioneering example is Znn.com~\cite{znn} that provides a webserver system of a simplified news site. Znn.com's testing environment simulates the slash-dot effect, which are periods of abnormally high traffic that overload the system. Another example artifact is TAS~\cite{7194661}, which provides a service-based system, where services offered by third-party providers are dynamically composed into workflows delivering complex functionality under uncertainties such as changing workload and fluctuations in the network. DARTSim~\cite{8787123}, instead, offers an artifact for the evaluation and comparison of adaptation approaches for smart cyber-physical systems that are subject to sensing errors among other uncertainties. Yet another example artifact is RoboMax~\cite{RoboMax}, which provides an extensible repository of robotic mission adaptation exemplars. A recent study highlighted an increasing use of artifacts in the evaluation of research results, although the adoption by teams other than the developing team remains relatively low~\cite{MScthesis}. 

A recent analysis of the maturity of the field of self-adaptation~\cite{weyns2021introduction} showed that we are currently in the phases of internal and external enhancement and exploration according to the model of Redwine and Riddle~\cite{319568.319624}. Hence, the industrial application of self-adaptation in real-life practical and commercial applications is of crucial importance to reach full maturity. Yet, it is currently not clear to what extent the existing artifacts support research activities of problems in self-adaptation that are relevant to industry. 

To that end, this paper provides a set of guidelines for future artifact providers that aim at supporting industry-relevant research on self-adaptation. In addition, the guidelines enable assessing existing artifacts on their relevance to support industry-driven research. 
The guidelines are grounded on the data obtained from a large-scale survey with practitioners. 
The guidelines have been derived from the outcome of a series of working sessions during the in-person day at the 17th International Symposium on Software Engineering for Adaptive and Self-Managing Systems, SEAMS 2022.\footnote{\url{https://conf.researchr.org/track/seams-2022/seams-2022-papers}} Note that alignment of research with industry has been topic of study in other domains, for instance in control~\cite{SAMAD20201} or in computing systems in general, see e.g.,~\cite{SI2021,FCR}. 

The remainder of this paper is structured as follows. In Section~\ref{sect:background}, we present the process we followed to define the guidelines presented in this paper, and we summarize the data that was used as input for the working sessions. Section~\ref{sect:guidelines} then presents the guidelines. In Section~\ref{sect:templates}, we present a template to describe industry-relevant artifacts on self-adaptation. Finally, we conclude the paper in Section~\ref{sect:conclusion}.

\section{Background}\label{sect:background}

In this section, we give an overview of the process we followed to create the guidelines for artifacts, together with a brief summary of the data used to define the guidelines.

\subsection{Process used to Derive the Guidelines} 

\begin{figure}[ht!]
    \centering
    \includegraphics[width=0.5\textwidth]{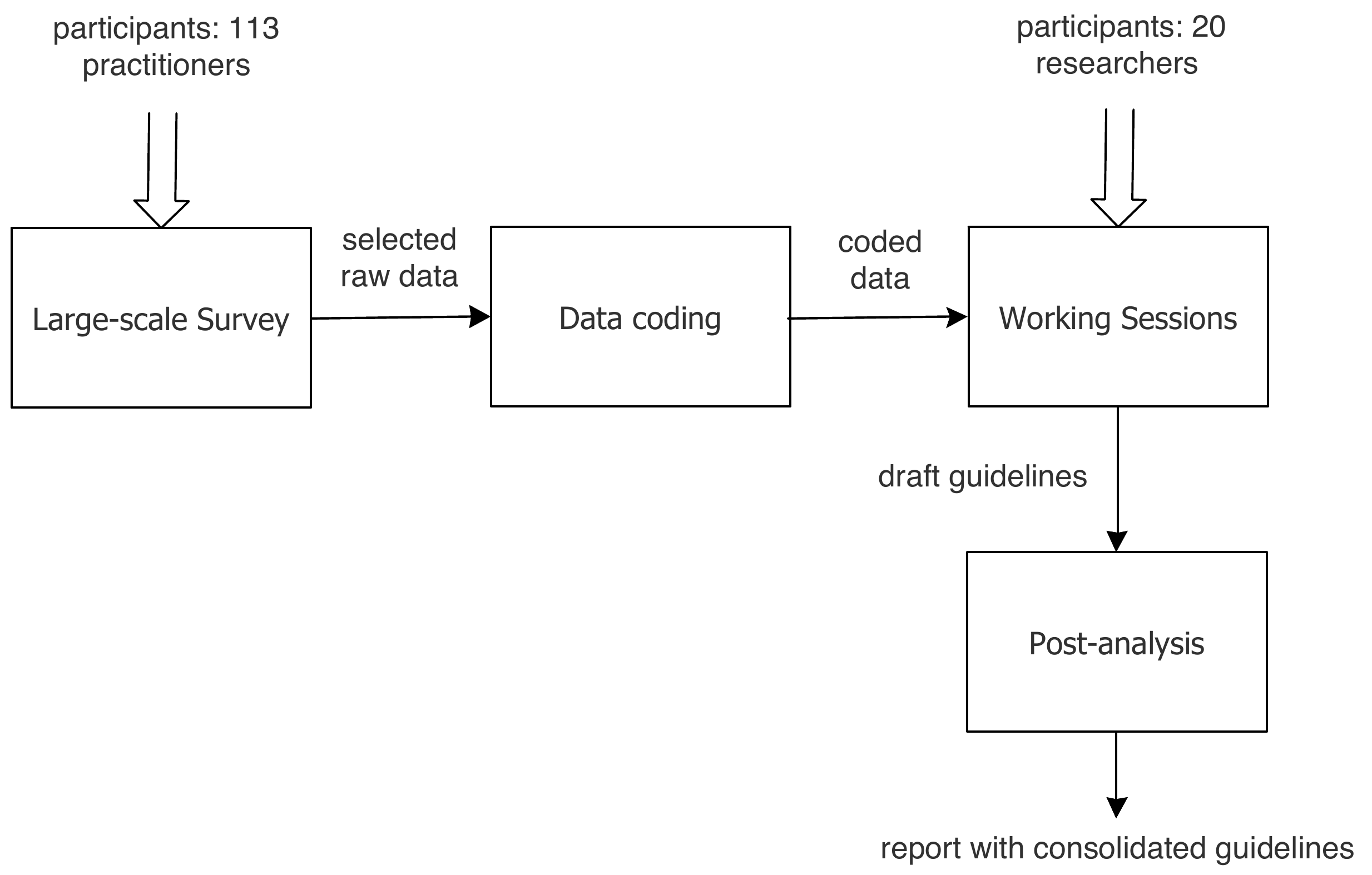}
    \caption{Process used to derive guidelines.}
    \label{fig:process}
\end{figure}

Figure~\ref{fig:process} shows the process we followed to derive the guidelines for artifacts. 

In a recent large-scale survey, we collected data from 113 practitioners\footnote{The participants range from different companies across the globe, active in a variety of domains; they have various roles in software engineering with a variety of experience (67\% have at least 9 years of experience)~\cite{arxiv.2204.06816}.} about: (a) the drivers and motivations to apply self-adaptation in industrial software, (b) the use cases and approaches used to realize self-adaptation, (c) the problems and challenges practitioners face, and (d) the opportunities practitioners see for self-adaptation. For a summary of the initial results, we refer to~\cite{arxiv.2204.06816}.
When defining the guidelines, we have used raw data on three specific topics: (a) difficulties and challenges practitioners face when applying self-adaptation, (b) problems for which they would appreciate support from researchers, and (c) opportunities for applying self-adaptation. In addition, we used categories and codes that were obtained from the raw data through inductive reasoning (i.e., moving from specific fragments in comments to general concepts)~\cite{Stol2016}. 

The coded data (elaborated below) was then used as input for the working sessions that were organized during the in-person part of SEAMS 2022. Concretely, we divided the group of 20 participants in three breakout groups. Each group drafted a set of guidelines during two sessions of 1.5 hours. Finally, the insights were brought together and discussed during a plenary session of one hour. After the event, we consolidated the guidelines and documented the results in this paper.

\subsection{Survey Data}

We summarize the results of the coding of the survey data by highlighting (a) difficulties and challenges practitioners face when applying self-adaptation, (b) problems for which they would appreciate support from researchers, and (c) opportunities for applying self-adaptation. 

Of all the surveyed practitioners, 35\% regularly face difficulties and challenges with applying self-adaptation, while 40\%  report that they sometimes face problems. Table~\ref{tab:codes_q4-2} shows the categories and codes collected from the answers. The table shows that practitioners face difficulties and challenges across different aspects of software engineering, from lifecycle issues to design, engineering processes,  and runtime.  

\begin{table}[hbt]
\caption{Difficulties and challenges in self-adaptive systems \\as perceived by practitioners}
\label{tab:codes_q4-2}
\begin{center}
\begin{tabular}{p{5cm}l}
\hline\noalign{\smallskip}
Categories and codes & \# \\
\noalign{\smallskip}\hline\noalign{\smallskip}

\rowcolor{lightgray} Lifecycle issues &  22 \\

Tuning/debugging & 10 \\

Limitations of tools/methods & 7 \\

System/environment evolution & 5  \\

\rowcolor{lightgray} Design challenges &  20  \\

Reliable/optimal design & 10 \\

Design complexity & 9 \\

Security & 1   \\

\rowcolor{lightgray} People and process issues & 14  \\

Skills/experience & 8  \\

Process and management & 5 \\

Documentation & 1 \\

\rowcolor{lightgray} Runtime challenges & 13  \\

Runtime uncertainty & 6  \\

Data collection/evaluation & 5 \\

Delayed/missing runtime changes & 2  \\

\noalign{\smallskip}\hline
\end{tabular}
\end{center}
\end{table}

Twenty percent of the practitioners express that they would frequently appreciate support from researchers to tackle problems they face, while 23\% would sometimes appreciate support. Table~\ref{tab:codes_q4-7} shows the categories and codes collected from the answers. The table shows that practitioners would appreciate support for tackling a variety of problems, from engineering to providing guarantees of trust, user interaction-related solutions, management of data, and the realization of a number of advanced features.

\begin{table}[hbt]
\caption{Problems for which practitioners would appreciate  support from researchers}
\label{tab:codes_q4-7}
\begin{center}
\begin{tabular}{p{3.6cm}l}
\hline\noalign{\smallskip}
Categories and codes & \# \\
\noalign{\smallskip}\hline\noalign{\smallskip}

\rowcolor{lightgray} Engineering & 22  \\
Architecture \& reuse & 7  \\
Adoption & 7 \\
Tools & 4  \\
Frameworks & 4  \\

\rowcolor{lightgray} Guarantees & 19 \\
Trustworthiness & 9  \\
Unknowns & 5   
 \\
Security & 5  \\

\rowcolor{lightgray} User interaction & 18  \\
User experience & 7 \\
Automation & 7 \\
User involvement & 4 \\

\rowcolor{lightgray} Data & 18  \\
Govern data & 8 \\
Assess data & 6  \\
Machine learning & 4 \\

\rowcolor{lightgray} Advanced features & 13  \\
Goals & 6  \\
Mechanisms & 4  \\
Testing & 3  \\

\noalign{\smallskip}\hline
\end{tabular}
\end{center}
\end{table}

Table~\ref{tab:codes_q4-9} shows the opportunities practitioners see for applying self-adaptation that are currently not exploited. The main opportunities are related to system activities and system properties. Other opportunities concern engineering activities and human involvement in the self-adaptation process.  

\begin{table}[hbt]
\caption{Opportunities for self-adaptation in industry that are currently not exploited as mentioned by practitioners}
\label{tab:codes_q4-9}
\begin{center}
\begin{tabular}{p{5cm}l}
\hline\noalign{\smallskip}
Categories and codes & \# \\
\noalign{\smallskip}\hline\noalign{\smallskip}

\rowcolor{lightgray} System activities & 26 \\
 Autonomous operation & 10 \\
 Autoscaling & 8 \\
 AutoML \& data management & 8 \\

\rowcolor{lightgray} System properties & 20  \\
Qualities & 12  \\
Cyber-security & 8  \\

\rowcolor{lightgray} Engineering activities & 15  \\
DevOps \& maintenance & 10 \\
Adaptation patterns \& libraries & 5  \\

\rowcolor{lightgray} Human involvement & 5 \\
Personalization & 4 \\
Human-machine interaction & 1 \\

\noalign{\smallskip}\hline
\end{tabular}
\end{center}
\end{table}

\section{Guidelines}\label{sect:guidelines}

We now present the consolidated guidelines. 
Table~\ref{tab:overview} provides an overview of the guidelines that are divided in three groups: required, recommended, and desirable guidelines. 
We discuss each group now in detail.  

\begin{table*}[h!]
\caption{Overview of guidelines for artifacts to support
industry-relevant research on self-adaptation}\smallskip\smallskip
\label{tab:overview}
\begin{center}
\begin{tabular}{l}
\hline\noalign{\smallskip}
\textbf{Required Guidelines}   \\\noalign{\smallskip}\hline\noalign{\smallskip}
 P1: Artifacts should take into account relevant surrounding elements \\
 \mbox{\ \ }P11: Artifact should take into account management of real-world data \\
  \mbox{\ \ }P12: Artifacts should take into account humans-on-the-loop \\
 \mbox{\ \ }P13: Artifacts should take into account industrial adoption of newly proposed solutions \\
 P2: Artifacts should take into account industry scale \\
 P3: Artifacts should provide industry-relevant metrics  \\
 P4: Artifacts should use an open policy  \\
 \noalign{\smallskip}\hline\noalign{\smallskip}
\textbf{Recommended Guidelines} \\\noalign{\smallskip}\hline\noalign{\smallskip}
R1: Integration with standard technologies \\
R2: Process-related artifacts \\
R3: Generality and extensibility of artifacts \\
R4: Artifacts that facilitate collaborations across
communities \\
 \noalign{\smallskip}\hline\noalign{\smallskip}
\textbf{Desirable Guidelines} \\\noalign{\smallskip}\hline\noalign{\smallskip}
D1: Artifacts may be built to exist in an ecosystem \\
D2: Artifacts may leverage open-source projects  \\
\noalign{\smallskip}\hline
\end{tabular}
\end{center}
\end{table*}

\subsection{Required Guidelines}

This first group of guidelines are required for industry-relevant artifacts in self-adaptation. Yet, which of these guidelines are relevant and apply depends on the type of artifact and its concrete purpose.   

\begin{quote}\vspace{-10pt}
    P1: \textit{Artifacts should take into account relevant surrounding elements} 
\end{quote}\vspace{-10pt}

Crucial for industry-relevant artifacts is that they take into account the context in which self-adaptation is applied. We refine this guideline in three concrete sub-guidelines. 

\begin{quote}\vspace{-10pt}
    P1.1: \textit{Artifact should take into account management of real-world data} 
\end{quote}\vspace{-10pt}

There are different kinds of data that a self-adaptive system must manage during operation.
Some are inherent to the functionality of the system, such as requests from users to a website and sensor readings in a cyber-physical system. Others are necessary for the self-adaptation process itself, such as monitoring data collected for analysis and planning.
It is important that artifacts can manage data with characteristics representative of industrial use, including volume, rate, noise, etc. Otherwise, research evaluated with artifacts may not be able to deal with the data needs of industrial use.

\begin{quote}\vspace{-10pt}
    P1.2: \textit{Artifacts should facilitate humans-on-the-loop}
\end{quote}\vspace{-10pt}

The initial research in engineering self-adaptive systems was mainly concerned with automating tasks~\cite{kephart2003vision,garlan2004rainbow,Weyns19}. In this research, the role of humans was in essence limited to providing high-level goals that the system should fulfill. Later, the role of humans-\textit{in}-the-loop became prominent~\cite{7194669}. In this perspective, humans can take different roles, from providing input to supporting decision-making and guiding adaptation actions. Yet, several researchers pointed out the potential implications and risks of uncertainty caused by humans-in-the-loop~\cite{esfahani2013usa,3487921}. 
On the other hand, practitioners point to the need for humans-\textit{on}-the-loop in self-adaptive systems. 
In industrial applications, humans should not be involved in the adaptation process all the time to avoid that they become a bottleneck. The preferred role of humans is instead supervision to achieve trustworthiness, avoid misuse or disuse of the automated system. In this regard, artifacts should consider different stakeholder roles: Users, operators, business owners, customers, etc., and emphasize their concerns such as feasibility, added value, and Return On Investment (ROI). 

\begin{quote}\vspace{-10pt}
    P1.3: \textit{Artifacts should take into account industrial adoption of newly proposed solutions} 
\end{quote}\vspace{-10pt}

Artifacts are developed to evaluate and demonstrate new research results, ultimately with the goal of advancing the state-of-practice. It is however well known that industry will only adopt research solutions if they are demonstrated to be effective in the context of real operational requirements. As such, transferring research results to the field requires organizations to conduct a preliminary evaluation (e.g., in the form of an industrial prototype, testbed or case study), which has a cost that may be seen as a risk by managers and decision-makers. Artifacts can play a major role in the reduction of such costs, thereby facilitating the experimentation of new solutions by industry and promoting the adoption of research results. In practice, artifacts should be built in such a way that allow industry to easily develop proofs of concept to test the effectiveness of new research solutions in their specific settings and adopt the results in their practice. 

\begin{quote}\vspace{-10pt}
    P2: \textit{Artifacts should take into account industry scale} 
\end{quote}\vspace{-10pt}

A representative scale is a key requirement for the industrial relevance of an artifact. Artifacts to support industry-relevant research should have a capacity and complexity that is characteristic of industrial practice. Taking into account real-world data as highlighted in guideline P1.1 is one specific facet of industry scale. Other aspects include the size of systems (e.g., a representative topology of an Internet of Things network),  requirements of systems (e.g., representative requirements of health care systems including privacy, performance, and cost), the load exercised on systems (e.g., realistic workloads used by Cloud resources), and underlying infrastructure (e.g., representative messaging techniques for micro-service architectures).   

\begin{quote}\vspace{-10pt}
\hspace{-5pt}P3: \textit{Artifacts should provide industry-relevant metrics} 
\end{quote}\vspace{-10pt}

To quantify the benefits and costs of self-adaptation, artifacts should provide industry-relevant metrics. On the one hand, such metrics should address quality characteristics of software systems to measure the satisfaction of quality requirements. This allows practitioners and researchers to evaluate a self-adaptive system built with an artifact from a technical perspective. On the other hand, such metrics should also address a business-oriented perspective, for instance, to evaluate a self-adaptive system built with an artifact economically (e.g., added value and ROI as also suggested by guideline P1.2). This allows customers and business owners to assess self-adaptation solutions developed with a specific artifact. Moreover, practitioners and researchers may select suitable artifacts based on their technical and business interest by taking these metrics into account. This allows them to easily pick up connection points for leveraging the self-adaptation mechanisms.

\begin{quote}\vspace{-10pt}
    P4: \textit{Artifacts should use an open policy} 
 \end{quote}\vspace{-10pt}

To facilitate adoption in industry, the artifacts need to (a) be openly available under clear licensing schemes, (b) leverage open standards (and preferably be based on standard technology, see guideline R1, and open-source projects, see guideline D2), and (c) ensure transparency. The transparency condition is crucial not only in terms of artifact trustworthiness, but also to support troubleshooting of defects in self-adaptation processes that are likely to be common in industry, given the reported shortage of expertise on the topic in the teams \cite{arxiv.2204.06816}, and might require further process-related support (see guideline R2). The transparency also facilitates repeatability (i.e., same team, same experimental setup, for the sake of validation), reproducibility (i.e., different team, same experimental setup, for the sake of statistical significance), and replicability (i.e., different team, different experimental setup, for the sake of wider adoption). As artifacts  may evolve over time, clear guidance on versioning and quality assurance is key to support industries willing to contribute to their extensions.

\subsection{Recommended Guidelines}

The second group of guidelines are recommended for industry-relevant artifacts in self-adaptation. We advise these guidelines as they will add to the relevance for industry. 

\begin{quote}\vspace{-10pt}
\hspace{-10pt}R1: \textit{\mbox{Integration with standard technology}} 
\end{quote}\vspace{-10pt}

Central to industrial relevance of artifacts is the use of standard technologies. This brings forward the following benefits:

\begin{itemize}
    \item \textit{Realism}: Artifacts developed with standard technologies allow for a closer level of realism due to the fact that the technologies used correspond to the actual technologies.\footnote{Besides technologies used, real-world data traces, realistic uncertainty profiles, etc. should be used to ensure industry relevance.}
    \item \textit{Engagement}: Using standard technologies ensures that a larger share of people is familiar with how the artifact functions. Examples of standard technologies (at the time of writing) include: Docker, Real-time Operating System (ROS) for Robotics, Carla (self-driving simulator), etc.
    \item \textit{Relevance}: The demonstrated benefits and evidence of techniques, solutions, or frameworks by using the artifact maps closer to industrial practice. More specifically, technology-specific settings of problems or opportunities put forward by industry connect more to the artifact when using the same technologies.
\end{itemize}

\begin{quote}\vspace{-10pt}
    R2: \textit{Process-related artifacts} 
\end{quote}\vspace{-10pt}

Process-related artifacts are those that are targeted at helping apply self-adaptation in industrial settings. Rather than tools that implement or support self-adaptation, these kinds of artifacts focus on software engineering processes (e.g., checklists) for designing, testing, and deploying them. For instance:

\begin{itemize}
    \item \textit{Design}: Guidance is needed on how to ensure that when applying self-adaptation in a particular context, that the system is complete and can provide a certain level of assurances. This requires that the right state is being modeled and monitored and the feedback loop is properly developed by combining tactics. E.g., it is not necessary to monitor elements that do not contribute to the decision-making about the system; similarly, if the self-adaptive system needs information in the planning or analysis process, this information needs to be stored in the knowledge that the self-adaptive system has about the system it is managing, and/or that appropriate monitoring is in place to retrieve that information in a timely and accurate way.
    \item \textit{Testing}: In order to practically apply self-adaptation, we need strategies for testing self-adaptive systems aligned with current practice, definitions of standard test metrics like coverage in the context of self-adaptation, and guidelines on how to provide a testing environment that drives the adaptations to be tested. Some of the research behind this has been done in~\cite{FC15,LC21,GW22}, but artifacts that can be used to operationalize self-adaptive testing would be helpful in guiding industry on how and what to apply.
    \item \textit{Deployment}: Once an adaptation decision is made, the system needs to enact the adaptation. This calls for strategies to apply adaptation actions in a safe way. Artifacts to integrate the execution of system adaptation aligned with state of the practice infrastructure for continuous deployment are key in helping practitioners with closing the feedback loop of self-adaptation. 
\end{itemize}

Process related artifacts may rely on, or take the form of best-practices documents, checklists for practitioners to use to ensure they deal with appropriate and common steps for self-adaptation, or test/deployment scripts that practitioners can use to manage CI/CD pipelines for self-adaptive systems.

\begin{quote}\vspace{-10pt}
    R3: \textit{Generality and extensibility of artifacts} 
\end{quote}\vspace{-10pt}

Generality and extensibility of artifacts add to their applicability and reusability. Therefore, it is important to define artifacts enabling their reuse both across application domains, such as robotics, automotive, and telecommunications, and/or axes of research/engineering activities, including requirements, architecture, verification, analysis, deployment, evolution, etc. Generality will leverage research results across domains and support the integration of research/engineering activities. Extensibility, on the other hand, enables the extension of existing artifacts to different domains and/or research/engineering activities. 
Note that this does not imply that domain-specific artifacts cannot not be effective and useful. For instance, RoboMax~\cite{RoboMax}, the artifact mentioned before, captures key sources of uncertainty and adaptation concerns of missions for the domain of robotic systems. Such domain-specific knowledge and know-how can add to the efficiency of applying an artifact.

\begin{quote}\vspace{-10pt}
    R4: \textit{Artifacts that facilitate collaborations across communities and with industry} 
\end{quote}\vspace{-10pt}

Industry requires simple and easily composable building blocks that can be shared for experimentation. The motivation for business gain could come from the composition of innovative ideas that have been developed by members of different scientific communities and mapped into robust workflows. Furthermore, science makes better progress when we have diversity of ideas and the people who generate the ideas~\cite{medin2012diversity}.
To enable cross-community collaborations and integration with industry, researchers should make it easy for others to build on top of the ideas and artifacts they generate. This calls for an infrastructure that easily facilitates automatic evaluation of methods/techniques/algorithms/etc. against benchmarks. Such an infrastructure will enable others to engage, while providing input and feedback from these diverse set of people. Several examples exist that have already demonstrated the added value; we give two examples:   
\begin{itemize}
    \item DLRM\footnote{\url{https://ai.facebook.com/blog/dlrm-an-advanced-open-source-deep-learning-recommendation-model/}} is a platform to which researchers from Computer Architecture and Systems and Recommendation Systems actively contribute. DLRM was developed by academics while placed as interns at Facebook. 
    \item MLCommons\footnote{\url{https://mlcommons.org/en/}} supported the development of multiple important benchmarks such as MLPerf and TinyML for researchers interested in machine learning for systems and systems for machine learning.  
\end{itemize}

In sum, we need an infrastructure, preferably open source, that makes it easier for researchers to share innovative ideas, for developers to experiment with ideas over a short time window (e.g., a weekend), and for software companies to make the ideas production-ready. This creates opportunities for more researchers, within and across different communities, to engage, and for industry to contribute to industry-relevant research efforts and engage with the research communities.

\subsection{Desirable Guidelines}

The third group of guidelines are desirable for industry-relevant artifacts in self-adaptation. These guidelines are worth having as they will add to the attractiveness for industry. 

\begin{quote}\vspace{-10pt}
    D1: \textit{Artifacts may be built to exist in an ecosystem} 
\end{quote}\vspace{-10pt}
Desirable properties of artifacts for industry-relevant research on self-adaptation are extensibility and integratability, so that they can accommodate surrounding elements (compare also guideline R3). 
In the past, the focus of self-adaptive systems has been on mechanisms to adapt systems' behaviors or structures, whereas surrounding elements (e.g., data management, human-machine interaction, and integration with basic infrastructure) have received little attention (see guideline P1). 
To integrate surrounding elements, preferable in a lightweight way, it is desirable not to focus on specifics of self-adaptive systems in isolation, but to design artifacts to co-exist within an ecosystem.
To support this guideline, for example, data management solutions may be provided as reusable components.
Moreover, artifacts should be designed so that they can be extended and integrated with other systems (see also guideline R4). This will add to a holistic perspective, which is a key aspect for industrial relevance of artifacts. 

\begin{quote}\vspace{-10pt}
    D2: \textit{Artifacts may leverage open-source projects}
\end{quote}\vspace{-10pt}

\begin{table*}[t!]
\caption{Proposed template for industry-relevant artifacts to support research on self-adaptive systems}\smallskip\smallskip
\label{tab:template}
\begin{center}
\begin{tabular}{p{3cm}p{13cm}}
\hline\noalign{\smallskip}
Element & Brief description \\
\noalign{\smallskip}\hline\noalign{\smallskip}

Problem & A brief description of the industry-relevant problem tackled by the artifact \\
Prerequisites & Requirements or constraints for using the artifact and the research results obtained with it \\
Technologies & Relevant state-of-the-practice technologies associated with the artifact \\
Outcome & The expected outcome of using the artifact and adopting the solution in practice (e.g., quantifiable by metrics) \\
Adoption & The relevant aspects required to adopt the research results obtained by using the artifact \\ 
\noalign{\smallskip}\hline\noalign{\smallskip}
Example & An exemplary solution to the problem and how it was adopted  \\
Related artifacts & Artifacts that address similar problems or that can be combined to enrich the solution  \\
\noalign{\smallskip}\hline\noalign{\smallskip}
Maintenance & A version history, and a reference to the maintainer of the artifact who keeps it up-to-date and responds to change requests  \\  
\noalign{\smallskip}\hline\noalign{\smallskip}
\end{tabular}
\end{center}
\end{table*}

To increase the industrial relevance of artifacts and their acceptance by industry, they would benefit from relying on well-known and robust projects. The use of open source projects may help artifacts to have a more stable underlying infrastructure. On the one hand, relying on existing infrastructure from open-source projects (e.g., containers, continuous integration tools, frameworks) reduces the learning curve for practitioners, and eases deployment and integration of the artifacts solutions with their technology base. On the other hand, using existing projects as a base, can lead artifacts to reach a readiness level accepted by industry faster.
Finally, we note that adopting an open-source process for the development of artifacts can be beneficial for their evolution and customization. As artifacts become more accepted, particular industries can customize a base artifact for their particular needs. As the artifacts evolve, new features can be integrated easily with branched and customized development.

\section{Template for Industry-Relevant Artifacts}\label{sect:templates}

To support industry-relevant research on self-adaptation, artifact descriptions would benefit from a standard structure, similarly to the uniform descriptions of design patterns. Standard template-based descriptions support both researchers and practitioners in searching, identifying, understanding, and comparing artifacts that are relevant for a given problem. Otherwise, the artifacts themselves have to be investigated, analyzed, and even executed to assess their relevance for the given problem and technology, which is a difficult and time-consuming effort. 

Table~\ref{tab:template} shows the proposed template for industry-relevant artifacts. 

The elements of the template were derived from the consolidated guidelines. The top part considers the basic elements to describe an artifact, from a description of the problem to the adoption of research results obtained with the artifact. The elements of the middle part provide an example solution and related artifacts that put the artifact in a broader context. Finally, the bottom part captures maintenance of the artifact in face of the rapid progress of technology in industry. Up-to-date maintained artifacts are essential to industrial relevance research. We anticipate that different types of templates might be needed to accommodate the description of different types of artifacts, such as testbeds, frameworks, datasets, and checklists. Finally, although the template elements proposed in Table~\ref{tab:template} were derived based on data obtained from practitioners in the context of self-adaptation, the elements of the template are generic and applicable to industry-relevant research in other areas. Yet, more study is required to underpin this observation. 

\section{Conclusions}\label{sect:conclusion}

Artifacts are key to systematic evaluation of new research results in a field of interest. We 
provided a set of empirically grounded guidelines for artifacts that aim at supporting industry-relevant research in the field of self-adaptation. The guidelines can be used by artifact providers for aligning future artifacts with industry needs as well as to assess the industrial relevance of existing artifacts. Finally, we proposed a template for describing industry-relevant artifacts to support research on self-adaptive systems. We hope that this paper will support the research community in self-adaptation to further enhance its maturity, the broader software engineering community to engage across their borders, and  practitioners in adopting research results to build better software systems.

\section*{Acknowledgment}

We are grateful to the practitioners that participated in the survey that we used as the basis for the presented guidelines. 

\bibliographystyle{plain}
\bibliography{biblio}

\end{document}